\documentclass[a4paper,11pt]{article}
\pdfoutput=1

\usepackage{amsmath,amssymb,hyperref}
\usepackage{fullpage}

\begin{document}

\title{Runaway directions in O'Raifeartaigh models}
\author{Zheng Sun\textsuperscript{*}, Xingyue Wei\textsuperscript{\dag}\\
        \normalsize\textit{Center for Theoretical Physics, College of Physical Science and Technology,}\\
        \normalsize\textit{Sichuan University, 29 Wangjiang Road, Chengdu 610064, P.~R.~China}\\
        \normalsize\textit{E-mail:}
        \textsuperscript{*}\texttt{sun\_ctp@scu.edu.cn,}
        \textsuperscript{\dag}\texttt{jiujfl@qq.com}
       }
\date{}
\maketitle

\begin{abstract}
R-symmetries, which are needed for supersymmetry (SUSY) breaking in O'Raifeartaigh models, often lead to SUSY runaway directions trough a complexified R-transformation.  Non-R symmetries also lead to runaway directions in a similar way.  This work investigates the occurrence of runaway directions of both SUSY and SUSY breaking types.  We clarify previous issues on fractional charges and genericness, and make a refined statement on conditions for runaway directions related to either R-symmetries or non-R symmetries.  We present a generic and anomaly-free model to show the existence of runaway directions related to non-R symmetries.  We also comment on the possibility to combine the non-R symmetry case to the R-symmetry case by an R-charge redefinition.
\end{abstract}

\section{Introduction}

In the study of supersymmetry (SUSY) model building~\cite{Martin:1997ns, Intriligator:2007cp}, R-symmetries are needed for SUSY breaking O'Raifeartaigh models because of the generic relations described by the Nelson-Seiberg theorem and its extensions~\cite{Nelson:1993nf, Sun:2011fq, Kang:2012fn}.  The study of metastable SUSY breaking~\cite{Intriligator:2006dd} also takes advantage from approximate R-symmetries~\cite{Intriligator:2007py}.  In addition, the SUSY breaking sector often contains global (non-R) symmetries which may be broken or gauged in the complete model.  It has been found that SUSY runaway directions related to R-symmetries are common in these models~\cite{Ferretti:2007ec, Ferretti:2007rq, Byakti:2012np}.  So the SUSY breaking vacuum due to the Nelson-Seiberg theorem can tunnel to the runaway direction through a non-perturbative process~\cite{Coleman:1977py, Coleman:1980aw} and becomes metastable.  To build a phenomenologically plausible model, one needs to tune the superpotential to give a long lifetime for the metastable vacuum against tunneling, or calculate quantum corrections which may stabilize fields at finite values along the runaway direction.

Runaway directions are obtained by a complexified R-transformation which pushes some field values to the infinity and satisfies SUSY equations at the asymptotic limit.  Proofs in previous literature utilize a generic form of R-symmetric superpotentials.  Fields are either assumed to have integer R-charges~\cite{Ferretti:2007ec, Ferretti:2007rq}, or redefined to alter the expression of the superpotential~\cite{Byakti:2012np}.  In this work we will clarify several issues in previous literature.  We point out that fractional R-charges complicate the superpotential, and renormalizability often leads to a non-generic superpotential.  Both of these issues make parts of the previous proof invalid.  We also show that models with no negatively R-charged field do not give runaway directions related to R-symmetries.  So the occurrence of runaway directions is less often than the estimation in previous literature.  And we make a refined statement on conditions for runaway directions of both SUSY and SUSY breaking types.

The technique of complexified transformation can also be applied to non-R $U(1)$ symmetries~\cite{Azeyanagi:2012pc}.  With an argument similar to the R-symmetry one, it can be shown that there exist runaway directions related to non-R $U(1)$ symmetries.  Besides the example in~\cite{Azeyanagi:2012pc} which needs another $Z_N$ symmetry for genericness, we present a new model to show the existence of such a runaway direction, with a generic superpotential respecting both an R-symmetry and a non-R $U(1)$ symmetry, and an anomaly-free $U(1)$ charge assignment.  Although non-R $U(1)$ symmetries can be absorbed into R-symmetries by an R-charge redefinition~\cite{Komargodski:2009jf}, the non-R $U(1)$ argument provides a clear view on the occurrence of runaway directions in certain examples and may have phenomenological advantages.

This paper is organized as follows.  Section 2 reviews runaway directions from R-symmetries and clarify several issues in previous literature; Section 3 discusses runaway directions from non-R $U(1)$ symmetries using a similar technique as in the R-symmetry cases, presents a generic and anomaly-free model to show the existence of runaway directions related to non-R symmetries, and comments on the R-charge redefinition to absorb non-R symmetries; Section 4 summarizes results of this paper, lists the conditions for runaway directions, and discusses issues of non-genericness and D-term runaway directions which are not covered in our work.

\section{Runaway directions from R-symmetries}

In O'Raifeartaigh models, the need for R-symmetries comes from their importance for SUSY breaking, as described by the Nelson-Seiberg theorem.  Given a superpotential $W(\phi_i)$, the SUSY breaking scale can be described by the F-terms $F_i = \partial_i W$ at the vacuum.  So a solution to equations $\partial_i W = 0$ gives a SUSY vacuum, and a minimum of the scalar potential $V = \lvert \partial_i W \rvert^2$ with $\partial_i W \ne 0$ gives a SUSY breaking vacuum.  R-symmetries are a special type of $U(1)$ symmetries which do not commute with the supercharge, thus rotate the superpotential by a phase.  The original Nelson-Seiberg theorem~\cite{Nelson:1993nf} claims that, for SUSY breaking in a generic O'Raifeartaigh model, a necessary condition is to have an R-symmetric superpotential, and a sufficient condition is to have spontaneous R-symmetry breaking at the vacuum.  An improved proof~\cite{Kang:2012fn} shows that the necessary and sufficient condition for SUSY breaking is to have an R-symmetric superpotential and more R-charge $2$ fields than R-charge $0$ fields.  A small modification towards an approximate R-symmetry gives metastable SUSY breaking~\cite{Intriligator:2007py}.  These results provide general guidelines for model building with R-symmetries.

Apart from vacua at finite field values, the vacuum structure of the model may also contains runaway directions.  Along the runaway direction, F-terms keep decreasing and reach zero at the asymptotic limit where some field values move to infinity.  One may also get SUSY breaking runaway directions if some F-terms remain to be non-zero at the limit.  A special type of SUSY runaway directions are often found in models with R-symmetries, by applying the complexified R-transformation:
\begin{equation}
\phi_i \to e^{R(\phi_i) \alpha} \phi_i , \quad \alpha \in \mathbb{R} ,
\end{equation}
where $R(\phi_i)$ is the R-charge of $\phi_i$.  Notice that this is not a symmetry of the whole SUSY action because of kinetic terms.  Since the superpotential always has R-charge $2$, F-terms are also rescaled under the complexified R-transformation:
\begin{equation}
\partial_i W \to e^{(2 - R(\phi_i)) \alpha} \partial_i W , \quad \alpha \in \mathbb{R} .
\end{equation}
In other words, $\partial_i W$ scales like a field with R-charge $2 - R(\phi_i)$.  If one of equations $\partial_i W = 0$ is solved, rescaling field values with the complexified R-transformation still solves the equation.

To show the occurrence of runaway directions, SUSY equations are classified according to their F-term R-charges:
\begin{align}
\partial_i W &= 0, \quad R(\phi_i) > 2 \Rightarrow R(\partial_i W) < 0 , \label{eq:2-1} \\
\partial_i W &= 0, \quad R(\phi_i) = 2 \Rightarrow R(\partial_i W) = 0 , \label{eq:2-2} \\
\partial_i W &= 0, \quad R(\phi_i) < 2 \Rightarrow R(\partial_i W) > 0 . \label{eq:2-3}
\end{align}
In SUSY breaking models, these equations can not be solved simultaneously.  If one can just solve equations of type \eqref{eq:2-1} and \eqref{eq:2-2}, the complexified R-transformation with $\alpha \to - \infty$ satisfies equations of type \eqref{eq:2-3} at the limit and gives a runaway direction.  An example of such case is the Witten's runaway model~\cite{Witten:1981kv}:
\begin{equation}
W = f X + \frac{1}{2} \lambda X^2 \phi ,
\end{equation}
with the R-charge assignment
\begin{equation}
R(X) = 2 , \quad R(\phi) = - 2 .
\end{equation}
The resulting SUSY runaway direction is given as:
\begin{equation}
\phi = - \frac{f}{\lambda X} , \quad X \to 0 .
\end{equation}
Similarly, if one can solve equations of type \eqref{eq:2-2} and \eqref{eq:2-3}, the complexified R-transformation with $\alpha \to + \infty$ satisfies equations of type \eqref{eq:2-1} at the limit and gives a runaway direction.  An example of such case is the spontaneous R-symmetry breaking model~\cite{Shih:2007av}:
\begin{equation}
W = f X + \lambda X \phi_1 \phi_2 + m_1 \phi_1 \phi_3 + \frac{1}{2} m_2 \phi_2^2 , \label{eq:2-4}
\end{equation}
with the R-charge assignment
\begin{equation}
R(X) = 2 , \quad R(\phi_1) = - 1 , \quad R(\phi_2) = 1 , \quad R(\phi_3) = 3 .
\end{equation}
The resulting SUSY runaway direction is given as:
\begin{equation}
X = \frac{m_2 f}{\lambda^2 \phi_1^2} , \quad \phi_2 = - \frac{f}{\lambda \phi_1} , \quad \phi_3 = \frac{m_2 f^2}{m_1 \lambda^2 \phi_1^3} , \quad \phi_1 \to 0 .
\end{equation}
If equations of type \eqref{eq:2-2} can not be solved, one needs to find a minimum of the potential from R-charge $0$ F-terms:
\begin{equation}
V_0 = \sum_{R(\phi_i) = 2} \lvert \partial_i W \rvert^2 .
\end{equation}
The complexified R-transformation leaves these non-zero F-terms invariant.  One may get SUSY breaking runaway directions by just solving equations of either type \eqref{eq:2-1} or type \eqref{eq:2-3}, and asymptotically satisfying the other.  In all cases, there are more fields than equations to be solved before taking the limit, and runaway directions should exist for a generic $W$.

For models with all R-charges to be integers, one can go further to prove some stronger statements~\cite{Ferretti:2007ec}.  In such a model, if there is only one R-charge $2$ field, \eqref{eq:2-2} and \eqref{eq:2-3} can always be solved and a runaway direction exists if there are fields with R-charge other than $0$, $1$, $2$.  A similar statement can be made with more than one R-charge $2$ fields, by satisfying some mild conditions of R-charge counting.  These statements can not be generalized to models with fractional R-charges, such as the tree-level R-symmetry breaking models~\cite{Komargodski:2009jf, Carpenter:2008wi, Sun:2008va}.  We show here an example of superpotential
\begin{equation}
W = \lambda \phi_1 (\phi_4 \phi_5 - m^2) + \mu \phi_2 \phi_4 + \nu \phi_3 \phi_5 + a \phi_5^2 \phi_6 + b \phi_4 \phi_6 \phi_7 + \sigma \phi_7^3 ,
\end{equation}
with the R-charge assignment
\begin{equation}
\begin{gathered}
R(\phi_1) = 2 , \quad R(\phi_2) = 20 / 9 , \quad R(\phi_3) = 16 / 9 , \\
R(\phi_4) = - 2 / 9 , \quad R(\phi_5) = 2 / 9 , \quad R(\phi_6) = 14 / 9 , \quad R(\phi_7) = 2 / 3 .
\end{gathered}
\end{equation}
All R-charges are fixed by the form of superpotential and can not be redefined to integers.  Such fractional R-charges complicate the superpotential and invalidate the stronger statements made by~\cite{Ferretti:2007ec}.  And detailed calculation indeed gives SUSY breaking at the global minimum without any runaway direction~\cite{Sun:2008va}.

An alternative attempt to show the existence of runaway directions has been provided~\cite{Byakti:2012np} using the field redefinition introduced in the original proof of the Nelson-Seiberg theorem~\cite{Nelson:1993nf}:
\begin{equation}
W = \phi_1^{2 / R(\phi_1)} f(Y_2, \dotsc , Y_n) , \quad \text{with } Y_i = \phi_i / \phi_1^{R(\phi_i) / R(\phi_1)} . \label{eq:2-5}
\end{equation}
$\partial_{i \ge 2} W = 0$ equations can be satisfied by solving $\partial_{Y_i} f = 0$ for a generic function $f$.  And the rest equation $\partial_{\phi_1} W = 0$ can be satisfied asymptotically by applying a proper complexified R-transformation.  The asymptotic limit either gives a SUSY runaway direction if $R(\phi_1) \in (- \infty, 0) \cup (2, + \infty)$, or pushes $\phi_1$ to zero which is a singular point in the field redefinition.  The result sounds too strong to be true:  In contrast to the previous R-charge counting requirement, only one field with R-charge negative or greater than $2$ is needed for the runaway direction.  It turns out that the discrepancy comes from non-genericness of $f(Y_2, \dotsc , Y_n)$.  The field redefinition from $\phi_i$'s to $Y_i$'s introduces fractional and negative powers of $\phi_1$, which have to be combined to integer and positive powers in the expression of $W$.  A renormalizable $W$ can have only polynomial terms up to cubic.  So many combinations of $Y_i$'s, although allowed by the R-symmetry, can not show up in $f$.  Such non-genericness can be demonstrated in the spontaneous R-symmetry breaking model \eqref{eq:2-4}.  We identify the $\phi_1$ in \eqref{eq:2-4} as the $\phi_1$ in \eqref{eq:2-5}.  A term like $Y_2 Y_3$ in $f$ corresponds to $\phi_1^2 \phi_2 \phi_3$ in $W$ which respects the R-symmetry but is not renormalizable.  So we see renormalizability puts a strong constraint on the possible form of $f$, and makes this statement through redefined fields invalid.

Finally, we make a comment on the case with no negatively R-charged field.  Since $W$ has R-charge $2$, such models also contain no field with R-charge larger than $2$.  So any field in such a model has R-charge in the $[0, 2]$ range.  There is no equation of type \eqref{eq:2-1}.  If \eqref{eq:2-2} and \eqref{eq:2-3} both are solved, one gets a SUSY vacuum.  If only \eqref{eq:2-2} is solved, the needed complexified R-transformation with $\alpha \to - \infty$ to solve \eqref{eq:2-3} asymptotically pushes all positively R-charged fields to zero, and leaves R-invariant fields unchanged.  Unlike the proof in~\cite{Byakti:2012np}, no field redefinition has been done and $W$ remains to be a polynomial.  The asymptotic limit is actually a SUSY vacuum at finite field values.  So we have shown that at least one field with negative R-charge is necessary for the existence of runaway directions related to R-symmetries.

\section{Runaway directions from non-R $U(1)$ symmetries}

The technique of complexified transformation used in the previous proof can also be applied to non-R $U(1)$ symmetries~\cite{Azeyanagi:2012pc}.  The procedure is almost the same to what has been done in the R-symmetry case.  The complexified $U(1)$ transformation is
\begin{equation}
\phi_i \to e^{Q(\phi_i) \alpha} \phi_i , \quad \alpha \in \mathbb{R} ,
\end{equation}
where $Q(\phi_i)$ is the $U(1)$ charge of $\phi_i$.  Notice again that this is not a symmetry of the whole SUSY action because of kinetic terms.  The superpotential must be invariant under the symmetry, thus has $U(1)$ charge $0$, F-terms are also rescaled under the complexified $U(1)$ transformation:
\begin{equation}
\partial_i W \to e^{- Q(\phi_i) \alpha} \partial_i W , \quad \alpha \in \mathbb{R} .
\end{equation}
In other words, $\partial_i W$ scales like a field with $U(1)$ charge $- Q(\phi_i)$.  If one of equations $\partial_i W = 0$ is solved, rescaling field values with the complexified $U(1)$ transformation still solves the equation.  SUSY equations are classified according to their F-term $U(1)$ charges:
\begin{align}
\partial_i W &= 0, \quad Q(\phi_i) > 0 \Rightarrow Q(\partial_i W) < 0 , \label{eq:3-1} \\
\partial_i W &= 0, \quad Q(\phi_i) = 0 \Rightarrow Q(\partial_i W) = 0 , \label{eq:3-2} \\
\partial_i W &= 0, \quad Q(\phi_i) < 0 \Rightarrow Q(\partial_i W) > 0 . \label{eq:3-3}
\end{align}
In SUSY breaking models, these equations can not be solved simultaneously.  If one can just solve equations of type \eqref{eq:3-1} and \eqref{eq:3-2}, the complexified $U(1)$ transformation with $\alpha \to - \infty$ satisfies equations of type \eqref{eq:3-3} at the limit and gives a runaway direction.  Similarly, if one can solve equations of type \eqref{eq:3-2} and \eqref{eq:3-3}, the complexified $U(1)$ transformation with $\alpha \to + \infty$ satisfies equations of type \eqref{eq:3-1} at the limit and gives a runaway direction.  If equations of type \eqref{eq:3-2} can not be solved, one needs to find a minimum of the potential from $U(1)$ charge $0$ F-terms:
\begin{equation}
V_0 = \sum_{Q(\phi_i) = 0} \lvert \partial_i W \rvert^2 .
\end{equation}
The complexified $U(1)$ transformation leaves these non-zero F-terms invariant.  One may get SUSY breaking runaway directions by just solving equations of either type \eqref{eq:3-1} or type \eqref{eq:3-3}, and asymptotically satisfying the other.  In all cases, there are more fields than equations to be solved before taking the limit, and runaway directions should exist for a generic $W$.

The literature~\cite{Azeyanagi:2012pc} provides an example with the following superpotential
\begin{equation}
W = f X_1 + \lambda X_1 \phi_1 \phi_2 + m X_2 \phi_2 + \lambda' X_3 \phi_1 \phi_3 ,
\end{equation}
and the R-charge and $U(1)$ charge assignment
\begin{gather}
R(X_1) = R(X_2) = R(X_3) = 2 , \quad R(\phi_1) = R(\phi_2) = R(\phi_3) = 0 ,\\
Q(X_1) = Q(\phi_3) = 0 , \quad Q(X_2) = Q(\phi_1) = 1 , \quad Q(X_3) = Q(\phi_2) = - 1 .
\end{gather}
It is noticed that only the first three terms of $W$ are necessary for showing the runaway direction along the $\phi_1$ direction.  But a negatively charged field $X_3$ is needed for anomaly cancellation if the $U(1)$ is promoted to a gauge symmetry at high scale.  Another neutral field $\phi_3$ in the fourth term is needed to keep the runaway direction intact.  And an additional $Z_{N>2}$ symmetry, under which $X_3$ and $\phi_3$ have charges $1$ and $-1$, is needed for $W$ to be generic.

Here we will present a new model with both an R-symmetry and a non-R $U(1)$ symmetry, but its genericness does not rely on extra symmetries.  Although extra symmetries like the $Z_N$ in the previous model are generally accepted for realistic model building, our model provides an alternative way for genericness and simplicity.  The superpotential is
\begin{equation}
W = X_1 (f + \lambda \phi_1 \phi_2) + X_2 (g + \kappa \phi_1 \phi_3) + m X_3 \phi_2 , \label{eq:3-4}
\end{equation}
with the R-charge and $U(1)$ charge assignment
\begin{gather}
R(X_1) = R(X_2) = R(X_3) = 2 , \quad R(\phi_1) = R(\phi_2) = R(\phi_3) = 0 , \label{eq:3-5}\\
Q(X_1) = Q(X_2) = 0 , \quad Q(X_3) = Q(\phi_1) = 1 , \quad Q(\phi_2) = Q(\phi_3) = - 1 .
\end{gather}
The $U(1)$ charge assignment ensures the model to be anomaly-free when the $U(1)$ is gauged, so one can later use D-terms to stabilize fields at finite values along the runaway direction, as was done in~\cite{Azeyanagi:2012pc}.  Furthermore, the superpotential can be obtained from a more generic form
\begin{equation}
W = \tilde X_1 (\tilde f + \lambda \phi_1 \tilde \phi_2 + \mu \phi_1 \phi_3) + \tilde X_2 (\tilde g + \nu \phi_1 \tilde \phi_2 + \kappa \phi_1 \phi_3) + X_3 (m \tilde \phi_2 + m' \phi_3) ,
\end{equation}
which includes all terms respecting both symmetries.  A field redefinition
\begin{equation}
X_1 = \tilde X_1 + \frac{\nu}{\lambda} \tilde X_2 , \quad X_2 = (\frac{\mu}{\kappa} - \frac{\lambda m'}{\kappa m}) \tilde X_1 + (1 - \frac{\nu m'}{\kappa m}) \tilde X_2 , \quad \phi_2 = \tilde \phi_2 + \frac{m'}{m} \phi_3
\end{equation}
and a coefficient reassignment
\begin{equation}
f = \frac{\lambda}{\lambda \kappa - \mu \nu} ((\kappa - \frac{\nu m'}{m}) \tilde f + (\mu - \frac{\lambda m'}{m}) \tilde g) , \quad g = \frac{\kappa}{\lambda \kappa - \mu \nu} (- \nu \tilde f + \lambda \tilde g)
\end{equation}
converts $W$ to the form \eqref{eq:3-4}.  So we will regard the superpotential \eqref{eq:3-4} as generic.

Since there are equal number of R-charge $2$ and R-charge $0$ fields, SUSY vacua should generically exist.  But because the non-R $U(1)$ symmetry restricts the form of $W$, our model actually has no SUSY vacuum.  One can see this by working out the SUSY equations from \eqref{eq:3-4}:
\begin{gather}
\partial_{X_1} W = f + \lambda \phi_1 \phi_2 = 0 , \quad \partial_{X_2} W = g + \kappa \phi_1 \phi_3 = 0 , \quad \partial_{X_3} W = m \phi_2 = 0 , \label{eq:3-6} \\
\partial_{\phi_1} W = \lambda X_1 \phi_2 + \kappa X_2 \phi_3 = 0 , \quad \partial_{\phi_2} W = \lambda X_1 \phi_1 + m X_3 = 0 , \quad
\partial_{\phi_3} W = \kappa X_2 \phi_1 = 0. \label{eq:3-7}
\end{gather}
Equations \eqref{eq:3-7} can always be solved by setting all $X$'s to zero.  But equations \eqref{eq:3-6} can not be solved simultaneously.  Searching for stationary points of the scalar potential turns out that there is only a SUSY breaking saddle point with all $\phi$'s at zero.  Since all $X$ fields have R-charge $2$, one can not see the existence of runaway directions through a complexified R-transformation.  However, one can classify the unsolved equations according to their F-term $U(1)$ charges:
\begin{align}
\partial_{X_1} W &= f + \lambda \phi_1 \phi_2 = 0, \quad Q(X_1) = 0 \Rightarrow Q(\partial_{X_1} W) = 0 , \label{eq:3-8} \\
\partial_{X_2} W &= g + \kappa \phi_1 \phi_3 = 0, \quad Q(X_2) = 0 \Rightarrow Q(\partial_{X_2} W) = 0 , \label{eq:3-9} \\
\partial_{X_3} W &= m \phi_2 = 0, \quad Q(X_3) = 1 \Rightarrow Q(\partial_{X_3} W) = - 1 . \label{eq:3-10}
\end{align}
The runaway direction can be found by solving equations with $Q(X) \le 0$, i.e., equations \eqref{eq:3-8} and \eqref{eq:3-9}, and satisfying equation \eqref{eq:3-10} with the complexified $U(1)$ transformation at the $\alpha \to + \infty$ limit.  The resulting SUSY runaway direction is given as:
\begin{equation}
\phi_1 = - \frac{f}{\lambda \phi_2} , \quad \phi_3 = \frac{\lambda g \phi_2}{\kappa f} , \quad \phi_2 \to 0 , \quad X_1 = X_2 = X_3 = 0 . \label{eq:3-11}
\end{equation}

As shown in literature~\cite{Komargodski:2009jf}, for a choice of R-charges $r_i$'s and non-R $U(1)$ charges $q_i$'s, R-charges can be reassigned as $r'_i = r_i + a q_i$ for $a \in \mathbb{R}$.  One can try to absorb the $U(1)$ symmetry into the R-symmetry, for example, by taking $R'(\phi) = R(\phi) + Q(\phi)$.  The R-charge assignment \eqref{eq:3-5} becomes
\begin{equation}
R'(X_1) = R'(X_2) = 2 , \quad R'(X_3) = 3 , \quad R'(\phi_1) = 1 , \quad R'(\phi_2) = R'(\phi_3) = - 1 .
\end{equation}
The SUSY runaway direction \eqref{eq:3-11} can be described as a result of R-symmetry with this new R-charge assignment, by solving equations of type \eqref{eq:2-2} and \eqref{eq:2-3}, and satisfying equations of type \eqref{eq:2-1} with the complexified R-transformation at the $\alpha \to + \infty$ limit.  Although most runaway directions from non-R $U(1)$ symmetries have such an R-symmetry description, there may exist some special choice of R-symmetries and non-R symmetries which are preferred for phenomenology study.  For example, a certain choice of $U(1)$ may be gauged at high scale and provide the D-term~\cite{Azeyanagi:2012pc}.  And the non-R symmetry description provides a clear view on the occurrence of runaway directions in such models.

\section{Conclusion}

We have discussed two general types of runaway directions in O'Raifeartaigh models.  They are obtained by the technique of complexified transformation from R-symmetries and non-R $U(1)$ symmetries, both of which are common in SUSY model building.  After clarifying issues on fractional charges and genericness, the conditions for runaway directions of both SUSY and SUSY breaking types are given as follows.

For runaway directions related to R-symmetries, SUSY equations are classified to three types \eqref{eq:2-1}, \eqref{eq:2-2} and \eqref{eq:2-3} according to their F-term R-charges.  SUSY breaking means all equations can not be solved simultaneously.
\begin{enumerate}
\item If there is no negatively R-charged field, then there is no runaway direction related to R-symmetries.  At least one field with negative R-charge is necessary for the existence of runaway directions related to R-symmetries.
\item If equations of type \eqref{eq:2-1} and \eqref{eq:2-2} can be solved, complexified R-transformation can asymptotically satisfy equations of type \eqref{eq:2-3} and give a SUSY runaway direction.
\item If equations of type \eqref{eq:2-2} and \eqref{eq:2-3} can be solved, complexified R-transformation can asymptotically satisfy equations of type \eqref{eq:2-1} and give a SUSY runaway direction.
\item If equations of type \eqref{eq:2-2} can not be solved, one needs to find a minimum of the potential from R-charge $0$ F-terms.  Then solving either one of \eqref{eq:2-1} and \eqref{eq:2-3} and asymptotically satisfying the other leads to a SUSY breaking runaway direction.
\end{enumerate}

For runaway directions related to non-R $U(1)$ symmetries, SUSY equations are classified to three types \eqref{eq:3-1}, \eqref{eq:3-2} and \eqref{eq:3-3} according to their F-term $U(1)$ charges.  SUSY breaking means all equations can not be solved simultaneously.
\begin{enumerate}
\item If equations of type \eqref{eq:3-1} and \eqref{eq:3-2} can be solved, complexified $U(1)$ transformation can asymptotically satisfy equations of type \eqref{eq:3-3} and give a SUSY runaway direction.
\item If equations of type \eqref{eq:3-2} and \eqref{eq:3-3} can be solved, complexified $U(1)$ transformation can asymptotically satisfy equations of type \eqref{eq:3-1} and give a SUSY runaway direction.
\item If equations of type \eqref{eq:3-2} can not be solved, one needs to find a minimum of the potential from $U(1)$ charge $0$ F-terms.  Then solving either one of \eqref{eq:3-1} and \eqref{eq:3-3} and asymptotically satisfying the other leads to a SUSY breaking runaway direction.
\end{enumerate}

As an example, we have presented a model with both an R-symmetry and a non-R $U(1)$ symmetry.  The superpotential includes all generic terms respecting both symmetries, and the model is anomaly-free when the $U(1)$ is gauged.  One finds a runaway direction \eqref{eq:3-11} related to the non-R $U(1)$ symmetry and no runaway directions related to the R-symmetry.  The non-R $U(1)$ symmetry case can be combined to the R-symmetry case by an R-charge redefinition which absorbs the non-R symmetry.  But the non-R symmetry description provides a clear view on the occurrence of runaway directions in certain models and may have phenomenological advantages.

In all cases, there are more fields than equations to be solved before taking the asymptotic limit, and runaway directions should exist for a generic $W$.  In certain examples the specific form of $W$ may make such small number of equations still unsolvable, and one needs to check by detailed calculation whether SUSY or SUSY breaking runaway directions exist.  In addition, runaway directions may also appear after including D-terms~\cite{Intriligator:2007py, Witten:1981kv, Vaknin:2014fxa, Kobayashi:2017fgl} which are not discussed in our work, and provide various tools for model building.

\section*{Acknowledgement}

The authors thank Benrong Mu for helpful discussions. This work is supported by the National Natural Science Foundation of China under the grant number 11305110.

\end{document}